\documentclass[aps,prl,twocolumn,groupedaddress,amsmath,amssymb]{revtex4}

\usepackage{epsfig}
\usepackage{dcolumn}
\usepackage{bm}
\usepackage{color}

\definecolor{Red}{rgb}{1,0.2,0.2}
\definecolor{Orange}{rgb}{1,0.5,0}
\definecolor{Green}{rgb}{0.1,0.8,0.1}

\begin{document}

%%%%%%%%%%%%%% Begin Commands %%%%%%%%%%%%%%%%%%%%%%%%%%%%%%%%%%%%%%%%%%
\def\d{{\rm d}}
\def\eps{\varepsilon}
\def\mg{m_{\tilde{g}}}
\def\mba{m_{\tilde{b}_1}}
\def\mbb{m_{\tilde{b}_2}}
\newcommand{\fmslash}[1]{\displaystyle{\not}#1}
\def\lp{\left. }
\def\rp{\right. }
\def\lr{\left( }
\def\rr{\right) }
\def\le{\left[ }
\def\re{\right] }
\def\lg{\left\{ }
\def\rg{\right\} }
\def\lb{\left| }
\def\rb{\right| }
\def\beq{\begin{equation}}
\def\eeq{\end{equation}}
\def\bea{\begin{eqnarray}}
\def\eea{\end{eqnarray}}
%%%%%%%%%%%%%% End of Commands %%%%%%%%%%%%%%%%%%%%%%%%%%%%%%%%%%%%%%%%%

%%%%%%%%%%%% Begin Cover Page %%%%%%%%%%%%%%%%%%%%%%%%%%%%%%%%%%%%%%%%%%
\preprint{LPSC 12-076}
\preprint{MS-TP-12-03}
\title{LHC phenomenology of general SU(2)$\times$SU(2)$\times$U(1) models}
\author{Tom\'{a}\v{s} Je\v{z}o$^a$}
\email[]{jezo@lpsc.in2p3.fr}
\author{Michael Klasen$^b$}
\email[]{michael.klasen@uni-muenster.de}
\author{Ingo Schienbein$^a$}
\email[]{schien@lpsc.in2p3.fr}
\affiliation{$^a$ Laboratoire de Physique Subatomique et de Cosmologie,
 Universit\'e Joseph Fourier/CNRS-IN2P3/INPG,
 53 Avenue des Martyrs, F-38026 Grenoble, France\\
 $^b$ Institut f\"ur Theoretische Physik, Westf\"alische
 Wilhelms-Universit\"at M\"unster, Wilhelm-Klemm-Stra\ss{}e 9, D-48149
 M\"unster, Germany}
\date{\today}
\begin{abstract}
 General SU(2)$\times$SU(2)$\times$U(1) models represent a well-motivated
 intermediate step towards the unification of the Standard Model
 gauge groups. Based on a recent global analysis of low-energy and LEP
 constraints of these models, we perform numerical scans of their
 various signals at the LHC. We show that total cross sections for lepton
 and third-generation quark pairs, while experimentally easily accessible,
 provide individually only partial information about the model realized in
 Nature. In contrast, correlations of these cross sections in the neutral
 and charged current channels may well lead to a unique identification.
\end{abstract}
%     SUSY    ,had.coll,polarizd,sparticl
\maketitle
%%%%%%%%%%%% End of Cover Page %%%%%%%%%%%%%%%%%%%%%%%%%%%%%%%%%%%%%%%%%

\vspace*{-82mm}
\noindent LPSC 12-076 \\
\noindent MS-TP-12-03 \\
\vspace*{59mm}

%%%%%%%%%%%%%% Begin Section 1 %%%%%%%%%%%%%%%%%%%%%%%%%%%%%%%%%%%%%%%%%
\section{Introduction}

The Standard Model (SM) of particle physics is very successful in
describing a wealth of experimental data, but is widely believed to be
incomplete. One of the reasons is that it reposes on the {\em ad hoc}
gauge group SU(3)$_C\times$SU(2)$_L\times$U(1)$_Y$ with three unrelated
factors, of which the second one violates parity, while the third one
depends on the unphysical hypercharge.
Their unification in a larger, simple group is theoretically very attractive.
Beyond minimal SU(5), already ruled out from proton decay, the possible
unification groups have a rank larger than the SM and thus also contain
additional subgroups like U(1) or a second SU(2), 
which lead to 
%leading to
additional neutral and charged gauge bosons. 
These so-called $Z'$ and $W'$ bosons are
actively searched for at the Large Hadron Collider (LHC) at CERN. Importantly,
the discovery of a $W'$ boson naturally implies the existence of a $Z'$ boson, but
not vice versa.

%While the recent discovery of non-zero neutrino masses, possibly generated by
%a see-saw mechanism, and the prospect of parity restoration point in the
%direction of a left-right (LR) symmetric group containing a SU(2)$_R$, the
%larger masses of Majorana neutrinos or of third-generation vs.\
%first- and second-generation fermions hint at fermion un-unified (UU) or
%generation non-universal (NU) groups SU(2)$_2$ and SU(2)$_1$, broken at
%high and low (SM-like) vacuum expectation values (VEVs) $u$ and $v$.

While the recent discovery of non-zero neutrino masses, possibly generated by
a see-saw mechanism, and the prospect of parity restoration point in the
direction of a left-right (LR) symmetric group containing a SU(2)$_R$, 
the large hierarchy in the mass spectrum of the SM fermions motivates
fermion un-unified (UU) or generation non-universal (NU) groups SU(2)$_2$ and SU(2)$_1$, 
broken at high and low (SM-like) vacuum expectation values (VEVs) $u$ and $v$.
In general, a large variety of models with a second SU(2) subgroup exist.
Often denoted as G(221) models, 
they also appear naturally in larger unification groups like
SO(10) and E$_6$ and in many string theory compactifications 
and can be classified according to their symmetry breaking pattern 
\cite{Hsieh:2010zr}. Identification of SU(2)$_1$ with the one of the SM
implies in the simplest scenarios the breaking scheme SU(2)$_2\times$U(1)$_X
\!\!\to$U(1)$_Y$ at 
the scale
$u$
through a Higgs doublet ($D$) or triplet ($T$).
This scheme applies not only to LR \cite{Mohapatra:1974hk}, but also
leptophobic (LP), hadrophobic (HP) and fermiophobic (FP) models
\cite{Barger:1980ti}. In contrast, identification of U(1)$_X$
with the hypercharge group of the SM as in the UU \cite{Georgi:1989ic}
and NU \cite{Li:1981nk} models leads to the
breaking scheme SU(2)$_1\times$SU(2)$_2\!\!\to\,$SU(2)$_L$ through a bi-doublet
Higgs at the scale $u$ with the immediate consequence of $M_{Z'}^2/M_{W'}^2=
1+{\cal O}(v^2/u^2)$. 
Parameterizing the model dependence in terms of the tangent of the mixing angle
$\tan\phi=g_X/g_2$ (LR, LP, HP, FP) or $g_2/g_1$ (UU, NU) at the
first breaking stage, the ratio of the squared Higgs VEVs $x=u^2/v^2$, and the
alignment angle $\beta$ of the light Higgs fields in the case of the first
breaking pattern, Hsieh {\em et al.} have been able to perform a global analysis
of low-energy and electroweak precision data with an effective Lagrangian approach
\cite{Hsieh:2010zr}, which resulted essentially in lower bounds on the masses of the
$Z'$ and $W'$ bosons.

At the LHC, information about extended gauge symmetries can be obtained
from cross section measurements, {\em e.g.} of pairs of leptons \cite{Chiang:2011kq}
or top quarks \cite{Basso:2012sz} or their associated production with $W'$ bosons
\cite{Berger:2011xk}, or measurements of the top quark polarization
\cite{Gopalakrishna:2010xm,Berger:2011hn,Basso:2012sz}.
In this Letter, we demonstrate that total cross sections for lepton and third-generation
quark production, while easily accessible at the LHC, allow for a unique
identification of the correct G(221) model only when correlated among each other
in the neutral current (NC) and charged current (CC) channels. In the following
section, we summarize the theoretical assumptions and experimental constraints
that have entered our analysis. We then describe the setup of our
simulations and present our numerical results for the total LHC cross sections
and their correlations. Our conclusions are given in the last section.

%%%%%%%%%%%%%% Begin Section 2 %%%%%%%%%%%%%%%%%%%%%%%%%%%%%%%%%%%%%%%%%
\section{Theoretical assumptions and experimental constraints}

We assume that the LHC will operate after the 2013-2014 shutdown at its
design center-of-mass energy of $\sqrt{s}=14$ TeV and accumulate an
integrated luminosity of 10 to 100 fb$^{-1}$, so that cross sections down
to 10$^{-2}$ fb can be observed. 
In the Sequential Standard Model (SSM) \cite{Altarelli:1989ff},
which 
(although physically unmotivated) 
%(although physically unmotivated in the context of grand unified theories) 
is often taken as a benchmark scenario,
the $Z'$ and $W'$ bosons would then be accessible up to masses of 5 TeV
\cite{Ball:2007zza}, while with present ATLAS (CMS) data masses below 2.21 (1.94)
\cite{Collaboration:2011dca,cms-pas-exo-11-019} and 2.15 (2.27) TeV
\cite{Degenhardt:2011aw,cms-pas-exo-11-024} are excluded in the leptonic channels.
We assume one of these bosons to have been observed in at
least one channel and its mass to have been measured from the invariant mass
of a lepton ($e$, $\mu$), $t\bar{t}$, or $t\bar{b}$ pair or the Jacobian peak
of the transverse mass of a single lepton and missing transverse energy as
3.0 $\pm$ 0.1 TeV or 4.0 $\pm$ 0.1 TeV with a conservative error estimate
\cite{Ball:2007zza} \footnote{Estimating the mass uncertainty $\Delta M$ from the
resonance width $\Gamma$ and the number of events $N$ with $\Delta M=\Gamma/\sqrt{N}$
shows that in the G(221) models studied here the statistical error will often be smaller by up to a
factor of ten. It will then become comparable to the experimental resolution of currently
about  2\% and ultimately 0.5\%.}. Since the cross section times branching
ratio to electrons and muons in the SSM is about six times larger for $W'$ than
for $Z'$ bosons, the former may well be discovered first \cite{atl-phys-pub-2011-002},
but this order is not essential to our study.

Apart from the $W'$ or $Z'$ mass, we impose the global constraints mentioned
above \cite{Hsieh:2010zr}, which we parameterize in terms of $\tan\phi$ and the
mass of the other new gauge boson.
We translate these constraints through an interaction Lagrangian
\bea
 {\cal L}&=& {g_L\over\sqrt{2}}
 \le\bar{u}_i\gamma^\mu\lr (C_{q,L}^{W'})_{ij}P_L+(C_{q,R}^{W'})_{ij}P_R\rr d_j\rp\nonumber\\
 &&\lp
%\quad\
 +\bar{\nu}_i\gamma^\mu\lr (C_{l,L}^{W'})_{ij}P_L+(C_{l,R}^{W'})_{ij}P_R\rr e_j\re W'_\mu\nonumber\\
 &+& {g_L\over c_\theta}\,
 \le\sum_q \bar{q}_i\gamma^\mu\lr (C_{q,L}^{Z'})_{ij}P_L+(C_{q,R}^{Z'})_{ij}P_R\rr q_j\rp\nonumber\\
 &&\lp
%\quad\
 +\sum_l \bar{l}_i\gamma^\mu\lr (C_{l,L}^{Z'})_{ij}P_L+(C_{l,R}^{Z'})_{ij}P_R\rr l_j\re Z'_\mu
\eea
into bounds on the left- and right-handed coupling constants $C_{L,R}$,
where the sums extend over up- and down-type quark flavors and
over neutrinos and charged leptons, respectively.
Scanning over the allowed parameter space we then obtain predictions
for the lepton ($e$ or $\mu$) and
third-generation ($t$ or $b$) quark production cross sections.

The global
constraints have been obtained on the basis of a number of theoretical
assumptions, in particular generation-diagonal, perturbative gauge couplings
(smaller than $\sqrt{4\pi}$), minimal (doublet, triplet or bi-doublet) Higgs
sectors with a hierarchy of VEVs ($u\gg v$), validity of the
Appelquist-Carazzone decoupling theorem, and negligible influence of
additional fermions required, {\em e.g.}, for the cancellation of gauge
anomalies. Fixed reference observables were the electromagnetic fine
structure constant $\alpha$, the Fermi constant $G_F$ determining $v^2$,
and the mass of the SM $Z$ boson $M_Z$ fixing the tangent of the weak mixing
angle $s_\theta/c_\theta=g_Y/g_L$. Similarly to the SM analysis of the Particle
Data Group (PDG) \cite{Nakamura:2010zzi}, the three free parameters were fit to 37
observables, of which the
most important ones were the total hadronic cross section at the $Z$ pole,
the $b$-quark forward-backward asymmetry, the neutrino-nucleon deep
inelastic scattering cross section, and the parity-violating weak charge
of Caesium 133. Low-energy constraints like BR($b\to s\gamma$) requiring
information on the extended flavor structure of the models ({\em e.g.} the
right-handed CKM matrix) were voluntarily omitted. Fixing the top quark and
light Higgs-boson masses to their SM best-fit values had little influence
on the results, which led to lower bounds on the $Z'$ and $W'$ masses
ranging from 0.3 to 3.6 TeV depending on the particular G(221) model.

% Our assumptions:
% - CKM matrix = true
% - couplings taken to be real
% - $C_{q,L}^{Z'}$ flavor-diagonal
% - right-handed neutrinos appear in LR and HP models, properties unknown, sigma_lnu not studied
% - right-handed CKM matrix needed in LR and LP models, taken same as left

%%%%%%%%%%%%%% Begin Section 3 %%%%%%%%%%%%%%%%%%%%%%%%%%%%%%%%%%%%%%%%%
\section{LHC phenomenology}

Total cross sections for the production of $W'$ and $Z'$ bosons decaying
into leptons and third-generation quarks have been simulated with the
Monte Carlo program {\tt Pythia 6.4} \cite{Sjostrand:2006za}, which we
have supplemented by terms accounting for the interferences of new and SM
charged bosons. While the latter can have a significant influence on the shape of the
resonance region, their impact on the total cross sections is small.
% parton shower and hadronization not simulated, but effect also small
Monte Carlo generators including next-to-leading order (NLO) QCD corrections
exist, but only for leptonically decaying $Z'$ and $W'$ bosons
\cite{Fuks:2007gk,Papaefstathiou:2009sr}, so that we have not
made use of them in this study for consistency. Furthermore, these
corrections are modest (typically about 30\%) and largely
model-independent, so that they would shift all contours in the same
direction and not change our conclusions. The masses of the
$t$ and $b$ quarks and SM $Z$ and $W$ bosons are fixed to their
PDG values \cite{Nakamura:2010zzi}, and we use the CTEQ6
% 4.8 GeV, 175 GeV (depend on ren. scheme!), 91.188, 80.45
LO analysis of parton density functions (PDFs) \cite{Pumplin:2002vw}.
% $\alpha_s(M_Z)=0.1184$ does not enter (EW process, no parton showers)
For a study of the PDF uncertainties on SM and new weak gauge boson
production cross sections we refer the reader to Ref.\ \cite{Martin:2009iq}.

Following the ATLAS leptonic analyses \cite{Collaboration:2011dca,Degenhardt:2011aw},
we require electrons to have transverse energy $E_T>25$ GeV and lie within the rapidity
ranges $|\eta| < 1.37$ or $1.52 < |\eta| < 2.47$. 
%
%For leptonic CCs, a missing energy $E_T^{\rm miss}>25$
For $\ell \nu$ final states, a missing energy $E_T^{\rm miss}>25$
GeV is imposed. In the third-generation quark channels, the ATLAS collaboration
reconstruct 
%reconstructs 
jets with the anti-$k_T$ algorithm and a radius of $R=0.4$ and
require 
%requires 
them to have $E_T>20$ GeV and $|\eta|< 4.5$ \cite{atlasttbar}. 
The total cross
sections are then still largely dominated by SM backgrounds, simulated also
using {\tt Pythia 6.4}, which we suppress by
imposing an invariant or transverse mass larger than 75\% of the known new gauge
boson mass.
% and mass parameter 7.5 TeV, no effect
% QCD backgrounds not treated!

%
%%%%%%%%%%%%%% Begin Figure 1 %%%%%%%%%%%%%%%%%%%%%%%%%%%%%%%%%%%%%%%%%%
\begin{figure}
 \centering
 \epsfig{file=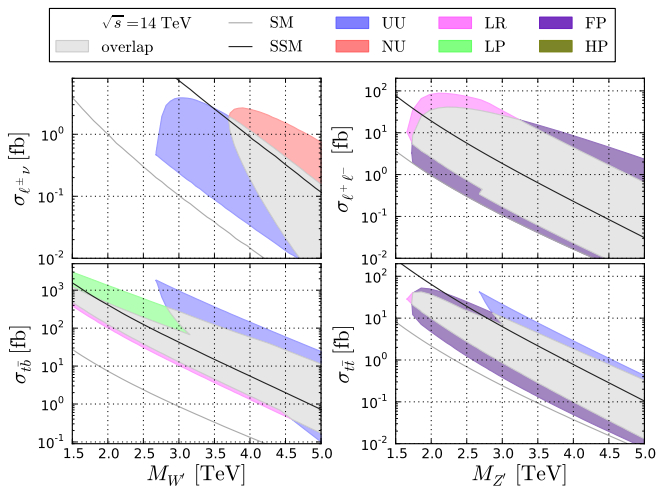,width=\columnwidth}
 \caption{\label{fig:1}Total CC (left) and NC (right) cross sections in the lepton (top) and third-generation quark (bottom) channels.}
% \emphNOTE{Shall we add a comment, that we assume $M_{W'}>1.5\ \text{TeV}$.}
\end{figure}
%%%%%%%%%%%%%% End of Figure 1 %%%%%%%%%%%%%%%%%%%%%%%%%%%%%%%%%%%%%%%%%
%

In Fig.\ \ref{fig:1} we show the resulting total CC (left) and NC (right) cross sections in the
lepton (top) and third-generation quark (bottom) channels. 
In all models, including the SSM shown for comparison, they are now 
by up to two orders of magnitude larger than the SM background
(which now depends on $M_{W',Z'}$ due to the transverse and invariant mass
cutoff, respectively),
but still 
exceed 10$^{-2}$ fb for masses up to 5 TeV. The shaded areas correspond to the
ranges in parameter space still allowed after the global analysis of low-energy and LEP
data. As one can observe, they exhibit a large overlap (grey). So
while the cross sections are experimentally easily accessible, the pre-LHC constraints
are clearly not strong enough to allow for an unambiguous identification of the gauge group
possibly realized in Nature. The same applies to the Higgs sector of the LR (pink), LP (green), HP
(purple), and FP (olive) models, for which regions with doublets and triplets are coded with
the same color, as they overlap almost completely. For these models, we make no predictions
for the lepton-neutrino channel, as we do not have any information about the right-handed
neutrino sector. Applying a $b$-tag to the quark channels would reduce their cross sections
by the corresponding efficiency of about 60\% \cite{atlasttbar}.

\begin{figure}
 \centering
 \epsfig{file=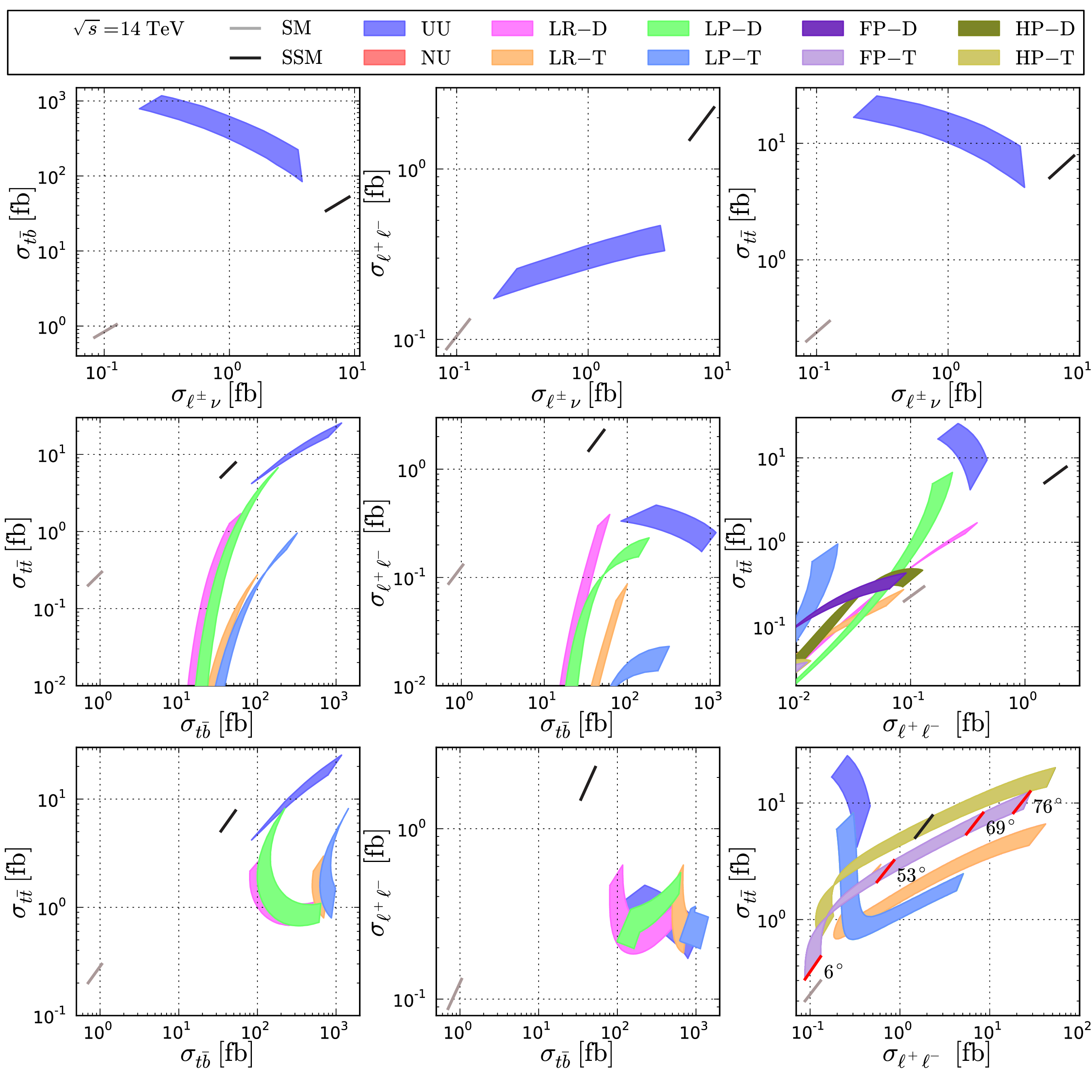,width=\columnwidth}
 \caption{\label{fig:2}Correlations of CC and NC lepton ($e$, $\mu$) and third-generation quark cross sections for fixed $M_{W'}=3.0\pm0.1$ TeV
 (top, center) and $M_{Z'}=3.0\pm0.1$ TeV (top, bottom).}
\end{figure}

In Fig.\ \ref{fig:2} we now correlate the total cross sections with each other assuming 
%
%%%%%%%%%%%%%% Begin Figure 2 %%%%%%%%%%%%%%%%%%%%%%%%%%%%%%%%%%%%%%%%%%
%%%%%%%%%%%%%% End of Figure 2 %%%%%%%%%%%%%%%%%%%%%%%%%%%%%%%%%%%%%%%%%
%
either a known mass of $M_{W'}=3.0\pm0.1$ TeV (top, center) or of $M_{Z'}=3.0\pm0.1$ TeV (top, bottom).
The shaded areas correspond now to the regions of parameter space allowed by the global fit and in
addition the uncertainty in the mass of the observed gauge boson. 
As already discussed in the introduction, for the UU and NU models $M_{W'}\simeq M_{Z'}$.
For the other models (LR, LP, FP, HP), we compute the mass that has not been fixed. 
More specifically, for the middle row $M_{Z'}$ is computed as a function of 
$M_{W'}$ and the other parameters at each point of the allowed parameter space \cite{Hsieh:2010zr}.
Similarly, for the bottom row $M_{W'}$ is determined as a function of $M_{Z'}$ and the other parameters.
Should the uncertainty be only 0.05 instead of 0.1 TeV, the width of these bands shrinks by more than 
a factor of two,
turning them almost into sharp lines.
% \emphNOTE{The previous sentence might be interpreted in the opposite sense: that is if the uncertainty doubles, the thickness extends by more than a factor of two -- we shall be careful here.}
As the plots in the first line always involve the lepton-neutrino channel and as masses below 3.6 TeV are
already excluded for the NU model \cite{Hsieh:2010zr} (cf.\ also Fig.\ \ref{fig:1}), we show there 
only predictions for the UU model.
%,in which $M_{W'}\simeq M_{Z'}$ (see above). 
With signal cross cross sections that are one to
three orders of magnitude larger than those of the SM and do not overlap with those of the
% (unphysical)
SSM
for which we also assume $M_{W'}=M_{Z'}$, 
it would therefore be easily identifiable. The plots in the other two lines do not involve the
lepton-neutrino channel, and we can therefore show predictions for all G(221) models except
for the excluded NU model and the HP and FP models, which are only accessible in the NC channels.
It is clear that the overlap is very much reduced
compared to Fig.\ \ref{fig:1}, making a unique identification of the underlying
gauge group possible in almost all cases. Even doublet (D) and triplet (T) Higgs fields can now be distinguished,
but this requires the observation of at least one CC channel and/or knowledge of the $W'$ mass.
The NC channels are obviously most useful if the $Z'$ mass is known (bottom right) and exhibit then
already by themselves a large discriminatory power of the gauge group. It would in particular be
sufficient to measure the $t\bar{t}$ and $\ell^+\ell^-$ cross sections with a precision of about 30\%.
%
%%%%%%%%%%%%%% Begin Figure 3 %%%%%%%%%%%%%%%%%%%%%%%%%%%%%%%%%%%%%%%%%%
\begin{figure}
 \centering
 \epsfig{file=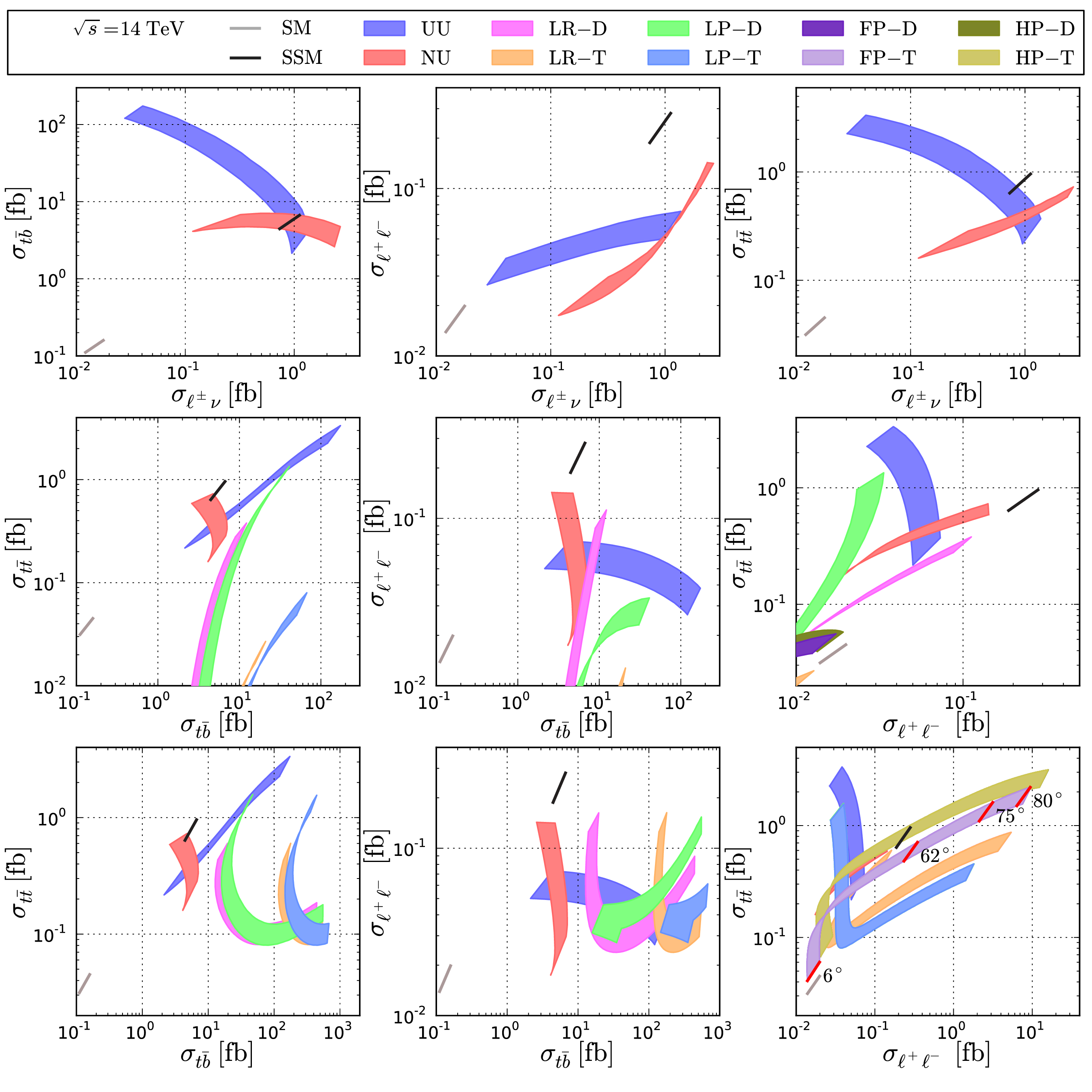,width=\columnwidth}
 \caption{\label{fig:3}Correlations of CC and NC lepton and third-generation quark cross sections for fixed $M_{W'}=4.0\pm0.1$ TeV
 (top, center) and $M_{Z'}=4.0\pm0.1$ TeV (top, bottom).}
\end{figure}
%%%%%%%%%%%%%% End of Figure 3 %%%%%%%%%%%%%%%%%%%%%%%%%%%%%%%%%%%%%%%%%
%
The position of these measurements in the correlation plane would then also give information on
the remaining model parameters, in particular
the mixing angle $\phi$ at the first breaking stage (cf.\ the red iso-lines in the bottom-right
plots of Figs.\ \ref{fig:2} and \ref{fig:3}).

A similar analysis is performed in Fig.\ \ref{fig:3}, where the $W'$ (top, center) and $Z'$ (top, bottom) mass
is now assumed to be $4.0\pm0.1$ TeV, respectively. This larger mass naturally leads to cross sections that are reduced
by about one order of magnitude with respect to those in Fig.\ \ref{fig:2}, but they remain observable and larger than the SM background.
For this higher mass, the NU model is no longer excluded by the global analysis and therefore added to the plots.
Correlating the lepton-neutrino with the $t\bar{b}$ and/or NC channels (top) makes it then clearly distinct
from the UU model. Since the third generation couples differently from the other two in the NU model, its
predictions for third-generation quark cross sections lead to almost constant lines or point-like
areas when correlated with the lepton cross sections or each other.

Conversely,
agreement of the experimental measurements with the SM predictions in
Figs.\ \ref{fig:1}-\ref{fig:3} would allow to considerably constrain
the parameter space of the different G(221) models or even to exclude
the corresponding new gauge boson masses altogether.

\begin{figure}
 \centering
 \epsfig{file=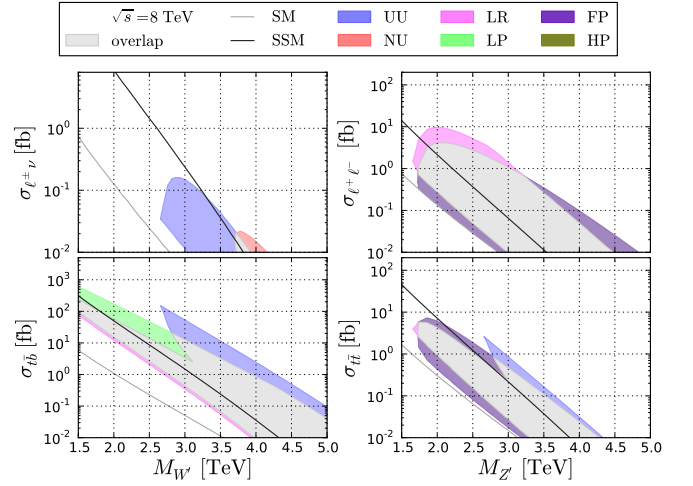,width=\columnwidth}
 \caption{\label{fig:4}Same as Fig.~\protect\ref{fig:1} for LHC at $\sqrt{s} = 8$ TeV.}
\end{figure}

\begin{figure}
 \centering
 \epsfig{file=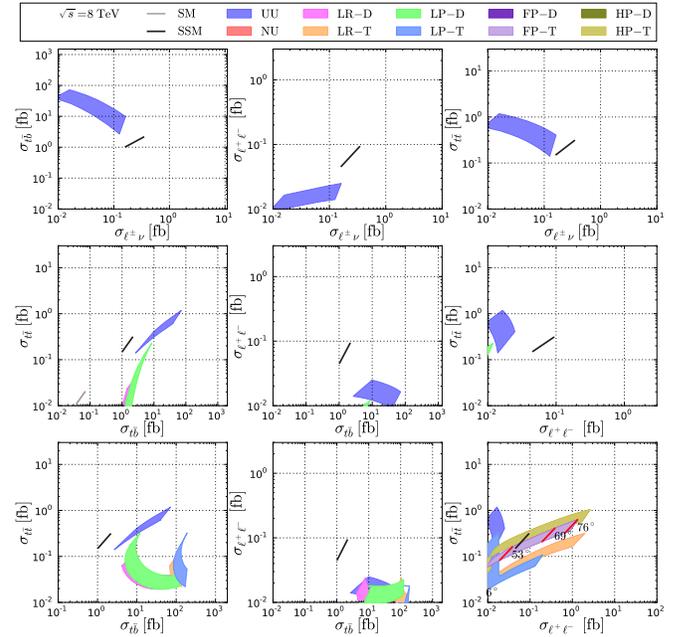,width=\columnwidth}
 \caption{\label{fig:5}Same as Fig.~\protect\ref{fig:2} for LHC at $\sqrt{s} = 8$ TeV.}
\end{figure}

For completeness, we also show predictions for the LHC operating at $\sqrt{s}=8$ TeV
in Figs.\ \ref{fig:4} and \ref{fig:5}.
%
%%%%%%%%%%%%%% Begin Figure 4 %%%%%%%%%%%%%%%%%%%%%%%%%%%%%%%%%%%%%%%%%%
%%%%%%%%%%%%%% End of Figure 4 %%%%%%%%%%%%%%%%%%%%%%%%%%%%%%%%%%%%%%%%%
%
%
%%%%%%%%%%%%%% Begin Figure 5 %%%%%%%%%%%%%%%%%%%%%%%%%%%%%%%%%%%%%%%%%%
%%%%%%%%%%%%%% End of Figure 5 %%%%%%%%%%%%%%%%%%%%%%%%%%%%%%%%%%%%%%%%%
%%
If one extrapolates the current LHC luminosity performance until the end of 2012, one
may estimate 60-80 fb$^{-1}$ so that measuring cross sections down to 10$^{-2}$ fb is not
unrealistic. 
The cross sections for resonances with a mass of 4 TeV turn out to be too small
and we therefore fix the mass to $3 \pm 0.1$ TeV in Fig.\ \ref{fig:5}.
The NU model can thus not be identified with data taken at 8 TeV.
Inspecting Fig.\ \ref{fig:5}, new gauge bosons with a mass of 3 TeV in the UU model 
could already be distinguished from the SSM if one of the third generation quark channels
is measurable (top). 
If both the neutral and charged current channels are measurable, one can
also distinguish it from the LR-D and LP-D models without making any assumptions on the
right-handed neutrino (center).
Distinguishing different ``doublet'' (LR-D, LP-D, \ldots) and ``triplet'' (LR-T, LP-T, \ldots) models 
is partly possible using the ($\sigma_{t\bar{t}},\sigma_{t\bar{b}}$) correlation (bottom left).
Furthermore, depending on the precision of the $Z'$ mass measurement,
it might also be feasible when correlating the neutral current lepton channel with the $t\bar{t}$ 
channel (bottom right) and then even lead to a first measurement of the mixing angle $\phi$.

%%%%%%%%%%%%%% Begin Section 4 %%%%%%%%%%%%%%%%%%%%%%%%%%%%%%%%%%%%%%%%%
\section{Conclusions}

In conclusion, we have proposed a novel and powerful method to distinguish general 
SU(2)$\times$SU(2)$\times$U(1) models, motivated experimentally, {\em e.g.}, by the 
observation of neutrino masses and theoretically as an intermediate
step towards the grand unification of the SM gauge groups. 
The total cross sections of the predicted charged and neutral
gauge bosons decaying into leptons and third-generation quarks were confirmed to be accessible at the LHC up to masses
of 5 TeV within the range of parameters allowed by a recent global analysis of low-energy and LEP constraints. Individually,
these cross sections did, however,
not allow for the unique identification of the underlying G(221) model. With a Monte Carlo simulation
and after applying realistic experimental cuts, we demonstrated that this does become possible
by correlating the charged and neutral current cross sections of leptons and third-generation quarks,
assuming only that the mass of either the $W'$ or the $Z'$ boson has been measured with a conservative uncertainty.
The mixing angle of the high-energy symmetry breaking stage will then also become measurable.
The correlations of two observables work nicely here because we
only have two (three) free parameters describing all the G(221) models.
In the general case of models with more parameters the identification
of suitable subsets of correlated observables would represent a first
step towards a global analysis of the parameter space.

\section*{Note added:}
After publication of this article four related articles have appeared discussing
the LHC phenomenology of models with an additional SU(2) group \cite{Chakrabortty:2012pp,Cao:2012ng,Du:2012vh,Abe:2012fb}.

%%%%%%%%%%%%%% Begin Section 4 %%%%%%%%%%%%%%%%%%%%%%%%%%%%%%%%%%%%%%%%%
\section*{Acknowledgments}

We thank R.\ Bonciani for his contributions during the initial stages
of this work and B.\ Clement, B.\ Dechenaux, P.A.\ Delsart, T.\ Junk, F.\ Ledroit, F.\ Lyonnet and A.\ Wingerter
for useful discussions.
This work has been supported by a Ph.D.\ fellowship of the French
Ministry for Education and Research and by the Theory-LHC-France
initiative of the CNRS/IN2P3.

%%%%%%%%%%%%%% Begin References %%%%%%%%%%%%%%%%%%%%%%%%%%%%%%%%%%%%%%%%

\end{document}